\documentclass[prd,aps,twocolumn,amsmath,amssymb,nofootinbib,preprintnumbers]
{revtex4}

\usepackage{graphicx}
\usepackage{dcolumn}
\usepackage{bm}
\usepackage{amsmath}
\usepackage{amsthm}

\usepackage{amsfonts}
\usepackage{euscript,bbm}
\usepackage{ifthen}
\usepackage{psfrag}
\usepackage{slashed}

\def\ls{\mathrel{\lower4pt\vbox{\lineskip=0pt\baselineskip=0pt
           \hbox{$<$}\hbox{$\sim$}}}}
\def\gs{\mathrel{\lower4pt\vbox{\lineskip=0pt\baselineskip=0pt
           \hbox{$>$}\hbox{$\sim$}}}}
\def\drawbox#1#2{\hrule height#2pt

\hbox{\vrule width#2pt height#1pt \kern#1pt
              \vrule width#2pt}
              \hrule height#2pt}

\def\Asym#1#2{\vcenter{\vbox{\drawbox{#1}{#2}
              \kern-#2pt       
              \drawbox{#1}{#2}}}}

\newcommand{\be}{\begin{equation}}
\newcommand{\ee}{\end{equation}}
\newcommand{\bea}{\begin{eqnarray}}
\newcommand{\eea}{\end{eqnarray}}

\newcommand{\neu}[1]{\ensuremath{\tilde{\chi}_{#1}^0}}

\newcommand{\chpm}[1]{\ensuremath{\tilde{\chi}_{#1}^{\pm}}}

\newcommand{\st}{\ensuremath{\tilde{t}}}

\newcommand{\gsim}{\lower.7ex\hbox{$\;\stackrel{\textstyle>}{\sim}\;$}}
\newcommand{\lsim}{\lower.7ex\hbox{$\;\stackrel{\textstyle<}{\sim}\;$}}

\newcommand{\tbar}{\overline{t}}

\newcommand{\ttbar}{t \bar{t}}

\newcommand{\met} {{E\!\!\!\!/_{\rm T}}}

\newcommand{\pT} {{p_{\rm T}}}

\newcommand{ \pythia } {{\tt PYTHIA}}

\newcommand{ \pgs }    {{\tt PGS4}}

\newcommand{ \madgraph } {{\tt MADGRAPH5}}

\begin{document}

%
\title{Probing Compressed Top Squarks at the LHC at 14 TeV}


\author{Bhaskar Dutta$^{1}$}
\author{Will Flanagan$^{1}$}
\author{Alfredo Gurrola$^{2}$}
\author{Will Johns$^{2}$}
\author{Teruki Kamon$^{1,3}$}
\author{Paul Sheldon$^{2}$}
\author{Kuver Sinha$^{4}$}
\author{Kechen Wang$^{1}$}
\author{Sean Wu$^{1}$}

\affiliation{$^{1}$~Mitchell Institute for Fundamental Physics and Astronomy, \\
Department of Physics and Astronomy, Texas A\&M University, College Station, TX 77843-4242, USA \\
$^{2}$~Department of Physics and Astronomy, Vanderbilt University, Nashville, TN, 37235, USA \\
$^{3}$~Department of Physics, Kyungpook National University, Daegu 702-701, South Korea \\
$^{4}$~Department of Physics, Syracuse University, Syracuse, NY 13244, USA 
}

\begin{abstract}

A feasibility study  is presented for the search of the lightest top squark ($\st$) in a compressed scenario, where its mass is approximately equal to the sum of the masses of the top quark and the lightest neutralino $\neu{1}$ and there exists no limit from the current 8-TeV data or from the 14-TeV projections. The study is performed in the final state of two $b$-jets, one lepton, large missing transverse energy, and two energetic  jets with a  large separation in pseudo-rapidity, in opposite hemispheres, and with large dijet mass. The analysis shows that the LHC could probe compressed top squarks  mass $\sim 300$ GeV with an integrated luminosity of $300$ fb$^{-1}$ for two ($t$+$\tilde\chi^0_1$) and three body ($b$+$W$+$\tilde\chi^0_1$) final states arising from the stop decay at $5\sigma$ significance with no systematic uncertainty. After including the systematics,  the significance  for $m_{\st}$ = 200 GeV and $\Delta M=7$ GeV is expected to be 6(3)$\sigma $ for 300 fb$^{-1}$ luminosity with 3(5)\% systematic uncertainty, while the significance becomes 4(2)$\sigma$ for  the same top squark mass with $\Delta M = -7$ GeV.



\end{abstract}
\maketitle


{\it {\bf Introduction-}} Weak-scale supersymmetry is a leading candidate for physics beyond the Standard Model (SM), as it addresses the hierarchy problem, gives gauge coupling unification, and (in $R$-parity conserving models) provides a robust dark matter (DM) candidate. 

The search for colored superpartners at the Large Hadron Collider (LHC) has so far yielded null results. The exclusion limits on squark ($\tilde{q}$) and gluino ($\tilde{g}$) masses, when they are comparable, are approximately $1.5$ TeV at $95\%$ CL with $20$ fb$^{-1}$ of integrated luminosity \cite{:2012rz, Aad:2012hm, :2012mfa, LHCsquarkgluino20ifb}. 

On the other hand, the bounds on the mass of the lightest top squark ($\st$) are less stringent. The vanilla scenario for $\st$ studies is to consider the direct QCD production of $\st$ pairs which decay to the top ($t$) and the lightest neutralino ($\neu{1}$) with $100 \%$ branching ratio. Exclusion limits in the $m_{\st}$-$m_{\neu{1}}$ plane have been obtained in this decay mode \cite{ATLASStop1, ATLASStop2}. 



The challenge of investigating $\st$ pair production lies in the huge background from top quark pair production. For this decay topology, the particles in the final state are identical to the $t \tbar$ background supplemented with missing transverse energy ($\met$). A number of analysis strategies have been proposed recently, covering both the fully hadronic \cite{hadronic} as well as the semi-leptonic \cite{semileptonic} final states. Search strategies in these final states using top taggers have also been pursued (for a review, see \cite{toptagger}). The projected top squark discovery mass reach and exclusion plots for the high-luminosity LHC have been studied by the ATLAS \cite{ATLASWhite} and CMS \cite{CMSWhite} Collaborations. We note as an aside that a couple of studies have focussed on other decay topologies, such as $\st \rightarrow b \chpm{1}$ \cite{Dutta:2013sta}, which are interesting from the point of view of well-tempered Bino/Higgsino or pure Higgsino dark matter.

The challenge is exacerbated when the mass gap between $\st$ and $t\,+\,\neu{1}$ is small. The $m_{\st} \, = \, m_{t} + m_{\neu{1}}$ line on the $m_{\st}$-$m_{\neu{1}}$ plane is a virtual Rubicon, and current exclusion bounds are non-existent near it. For $m_{\st} \sim 190 - 300$ GeV, the exclusion bounds come within $\Delta M = m_{\st}- (m_{t} + m_{\neu{1}}) \sim 15$ GeV. For $m_{\st} \sim 300 - 450$ GeV, there is significant degradation and exclusions only reach $\Delta M \, \sim \, 25$ GeV. For $m_{\st} > 450$ GeV, the smaller production cross-section leads to exclusion bounds with $\Delta M \gg 50$ GeV. The discovery reach at the 14 TeV LHC (LHC14) with 300 fb$^{-1}$ data, assuming an optimistic projection \footnote{The optimistic projection scales up $N_S$ and $N_{\rm{BG}}$ by "cross-section ratio times luminosity" ratio from 8-TeV analyses with its uncertainty reduced by 1/$\sqrt{r_{\rm{BG}}}$ or a minimum value of $10\%$.} of LHC8 results, is similar. In this \textbf{compressed} scenario, search strategies that rely on $\met$ to reduce $\ttbar$ background have poor performance. The challenge is even greater when $m_{\tilde{\chi}_{1}^{0}}$ becomes vanishingly small in the compressed region, so that $m_{\st} \sim m_{t}$. In this scenario, which is called the \textbf{stealthy} scenario ($\Delta M \, \sim \, 0$ GeV), the $\met$ discrimination between signal and background becomes very ineffective. These scenarios have been studied by several groups and the proposed strategies include a shape-based analysis of the $\met$ and $m_T$ distributions  \cite{Alves:2012ft}, rapidity gap and spin correlation observables \cite{Han:2012fw}, and optimized use of dileptonic $m_{T2}$ \cite{Kilic:2012kw}.

Similarly, probing the top squark in its three-body decay mode $\st \rightarrow bW \neu{1}$ is also difficult. The current exclusion limit on this mode at the 8 TeV LHC (LHC8) with $20$ fb$^{-1}$ of data from CMS starts from $m_{\st} = 200$ GeV, with $\Delta M =-25$ GeV.  The discovery reach at the 14 TeV LHC (LHC14) with 300 fb$^{-1}$ data is similar. For smaller $\Delta M$ there are no limits, e.g., the limit ceases to exist for $-150\leq\Delta M \leq -70$ GeV at $m_{\st}=200$ GeV. Although the current monojet searches can place contraints for $\Delta M<-150$GeV, those contraints are limited beyond $m_{\st}=200$ GeV.


The purpose of the current work is to propose search strategies for $\st$ pairs in the compressed scenario in the small $\Delta M$ region  using Vector Boson Fusion (VBF) tagging. VBF jet topologies have recently been proposed by the authors as a probe of the non-colored sector of supersymmetric models. Charged and neutral Wino production followed by decays to $\neu{1}$ via a light slepton has been studied in \cite{Dutta:2012xe,cho}, while VBF searches for Wino and Higgsino DM has been proposed in \cite{Delannoy:2013ata}. The classic mechanism for VBF searches in the case of noncolored particles occurs through the fusion of the W and Z weak bosons.  As shown in \cite{Dutta:2012xe} and \cite{Delannoy:2013ata}, the requirement of two energetic jets in the forward region with large dijet invariant mass is very effective in reducing SM backgrounds in the VBF analysis. The stops are produced  through gluon fusion, one of the dominant sources of production in the case of colored particles. However we still require two jets in the final state with large separation in pseudorapidity and large dijet invariant mass just like the VBF searches for pure electroweak production. We refer to this type of final state as VBF topology. 

 In contrast to other $\st$ searches where compressed spectra results in low $\met$, making it difficult to discriminate against $\ttbar$ background, VBF topologies naturally give rise to larger $\met$ since the momentum of the particles produced in the $\st$ system must balance the high $p_{T}$ of the scattered partons. Thus, in the compressed scenario, the $\neu{1}$ resulting from the $\st$ decay carries significant $\met$, providing better control of the $\ttbar$ background.






{\it {\bf Search Strategy -}} For this feasibility study, inclusive $\st \st^{*} \, + \,$ multi jets samples are generated with $\st$ masses in the range of $200$-$600$ GeV, keeping $\Delta M \sim \pm 7$ GeV. This region is not constrained by the present limit. There also exists no limit from the projections from the 14 TeV LHC. Both QCD and weak production processes of $\st \st^{*} \, + \,$ multi-jets are included. The $\neu{1}$ in our studies is mostly Bino, while the $\st$ is mostly $\st_R$ such that the dominant decay mode of the $\st$ is $\st \, \rightarrow t \neu{1}$ in the 2 body case and $\st \, \rightarrow t^{\ast} \neu{1}$ in the 3 body case. The signal in both $\Delta M \sim \pm 7$ GeV cases is 2 high $p_T$ jets + $2b\,+\,1 l+\met$. 
%
%
%
The other colored particles, neutralinos and charginos are assumed to be much heavier. 

Signal and background samples are generated with \madgraph \, \cite{Alwall:2011uj} followed by the parton showering and hadronization with \pythia \, \cite{Sjostrand:2006za} and the detector simulation using \pgs \, \cite{pgs}. 

We use pre-cut samples for signal and background to develop our search strategies. The pre-cut sample is is obtained using \madgraph run card level cuts. The search strategy is based on three steps. First, we use the unique features of VBF jet topology to reduce $V \, + \,$  jets backgrounds (where $V$ is either $W$ or $Z$). Second, we the use decay properties of the centrally produced $\st$ pair, namely the requirement of an isolated lepton and two $b$-tagged jets from a top quark, to further reduce light quark QCD backgrounds and other channels that are also produced by VBF topologies. Finally, the $\met$ distribution is used to assess the presence of a signal above the $t\bar{t}$ .

$(1)$ VBF cuts: the event is required to have a presence of at least two jets ($j_1$, $j_2$) satisfying: $(i)$ $\pT  \geq 75$ and $50$ GeV in $|\eta| \leq 4$; $(ii)$ $|\Delta \eta (j_1, j_2)| > 3.5$; $(iii)$ $\eta_{j_1} \cdot \eta_{j_2} < 0$; $(iv)$ dijet invariant mass $M_{j_1 j_2} \, > \, 500$ GeV.

$(2)$ One isolated lepton with $\pT \geq 20$ GeV and two loose $b$-jets with $\pT \geq 30$ GeV in $|\eta| < 2.5$ are required. The $b$-jet identification efficiency and fake rate are taken to be $70\%$ and $1\%$, respectively.

$(3)$ In order to highlight the effectiveness of the $\met$ distribution after the VBF topological selections, the cut flow tables with corresponding cross-sections at each stage are displayed under different considerations of the $\met$ phase space (e.g. $\met > 100$ GeV for $m_{\st} = 300$ GeV).



{\it {\bf Compressed Scenario -}}  The cut flow table with corresponding cross-sections at each stage are shown in Table \ref{tablestopbenchmark1} and \ref{tablestopbenchmark2} for  $\Delta M= \pm 7$ GeV. As mentioned, the $\met$ cuts are very effective  in improving the signal to background ratio.

\begin{table}[!htp]
\caption{Compressed scenario with $\Delta M=7$ GeV: Summary of the effective cross-sections (fb) for different benchmark signal points as well as the $\ttbar$ background at 14 TeV LHC. Masses and momenta are in GeV.}
\label{tablestopbenchmark1}
\begin{center}
\begin{tabular}{c c c c c}
\hline \hline

$(m_{\st}, m_{\neu{1}})$ & Selection &Signal &\,\, $\ttbar + $jets & S/B \\
\hline
\hline \\
$(200,20)$ & Pre cut & $5.4 \times 10^{4}$ & $6.9 \times 10^{5}$ & ---\\
$\Delta M=7$  & VBF & $1.8 \times 10^{3}$ & $3.8 \times 10^{4}$ & ---\\
& 1 lepton & 390 & $8.1 \times 10^{3}$ &--- \\
& 2 $b$-jets & 170 & $3.1 \times 10^{3}$ & $5.6 \times 10^{-2}$ \\
& $\met > 100$ & 44 & 680 &  $6.5 \times 10^{-2}$\\
\hline \\
$(300,120)$ & Pre cut & $7.4\times 10^{3}$ & $6.9 \times 10^{5}$ & ---\\
$\Delta M=7$ & VBF & 250 &  $3.8 \times 10^{4}$ &  ---\\
& 1 lepton & 56 & $8.1 \times 10^{3}$ & ---\\
& 2 $b$-jets & 32 & $3.1 \times 10^{3}$ & $1.0 \times 10^{-2}$ \\
& $\met > 100$ & 8.9 & 680 & $1.3 \times 10^{-2}$ \\
\hline \\
$(400,220)$ & Pre cut & $1.6\times 10^{3}$ & $6.9 \times 10^{5}$ & ---\\
$\Delta M=7$ & VBF & 62 & $3.8 \times 10^{4}$ &  ---\\
& 1 lepton & 14 & $8.1 \times 10^{3}$ &  ---\\
& 2 $b$-jets & 8.4 & $3.1 \times 10^{3}$ & $2.7 \times 10^{-3}$ \\
& $\met > 100$ & 4.8 & 680 & $7.0 \times 10^{-3}$\\
\hline \\
$(500,320)$ & Pre cut & 460 & $6.9 \times 10^{5}$ & ---\\
$\Delta M=7$ & VBF & 19 & $3.8 \times 10^{4}$ &  ---\\
& 1 lepton & 4.2 & $8.1 \times 10^{3}$ & ---\\
& 2 $b$-jets & 2.4 & $3.1 \times 10^{3}$ & $7.9 \times 10^{-4}$ \\
& $\met > 150$ & 1.5 & 250 & $6.0 \times 10^{-3}$ \\
\hline \hline

\end{tabular}
\end{center}
\end{table}




\begin{table}[!htp]
\caption{Compressed scenario with $\Delta M=-7$ GeV: Summary of the effective cross-sections (fb) for different benchmark signal points as well as the $\ttbar$ background at LHC14. Masses and momenta are in GeV.}
\label{tablestopbenchmark2}
\begin{center}
\begin{tabular}{c c c c c}
\hline \hline

$(m_{\st}, m_{\neu{1}})$ & Selection &Signal &\,\, $\ttbar + $jets & S/B\\
\hline
\hline\\ 
$(200,35)$ & Pre cut & $5.4 \times 10^{4}$ & $6.9 \times 10^{5}$ & ---\\
$\Delta M=-7$   & VBF & $1.4 \times 10^{4}$ & $3.8 \times 10^{4}$ &  ---\\
& 1 lepton & 270 & $8.1\times 10^{3}$ &  ---\\
& 2 $b$-jets & 79 & $3.1\times 10^{3}$ & $2.5\times 10^{-2}$ \\
& $\met > 100$ & 29 & 680 & $4.3\times 10^{-2}$\\
\hline \\
$(300,135)$ & Pre cut & $7.4 \times 10^{3}$ & $6.9 \times 10^{5}$ & ---\\
$\Delta M=-7$  & VBF & 220 & $3.8 \times 10^{4}$ &  ---\\
& 1 lepton & 43 & $8.1\times 10^{3}$ &  ---\\
& 2 $b$-jets & 12 & $3.1\times 10^{3}$ & $3.7 \times 10^{-3}$ \\
& $\met > 100$ & 6.7 & 680 & $9.8 \times 10^{-3}$\\
\hline \\
$(400,235)$ & Pre cut & $1.6 \times 10^{3}$ & $6.9 \times 10^{5}$ & ---\\
$\Delta M=-7$  & VBF & 51 & $3.8 \times 10^{4}$ &  ---\\
& 1 lepton & 10. & $8.1\times 10^{3}$ &  ---\\
& 2 $b$-jets & 2.8 & $3.1\times 10^{3}$ & $8.9\times 10^{-4}$ \\
& $\met > 200$ & 0.7 & 100 & $6.6 \times 10^{-3}$ \\
\hline \\

\end{tabular}
\end{center}
\end{table}

After all the cuts, the $t\bar{t}$ contribution is found to be the dominant background. The $W + $ jets as well as $WZ, WW$ events are expected to be negligible. The combined contribution from $W +$ jets, $WZ$, and $WW$ events is negligible. The requirement of the presence of the isolated lepton in the signal reduces the light flavor and gluon jets from QCD processes  effectively.

As $\Delta M$ increases, the $b$ jet becomes more energetic and the signal rate improves. In order to show this feature explicitly, let us choose  $m_{\st}$=300 GeV with  $m_{\neu{1}} = 150$ and 135 GeV. We find that after the $\met$ cut, the signal cross-sections are 5.0 fb and 6.7 fb for $m_{\neu{1}} = 135$ and 150 GeV, respectively. We note that the ability to identify soft b-jets ($p_{T}\sim 20$GeV) in a high pileup environment of the LHC14 is an important requirement for this analysis.

\begin{figure}[!htp]
\centering
\includegraphics[width=3.0in]{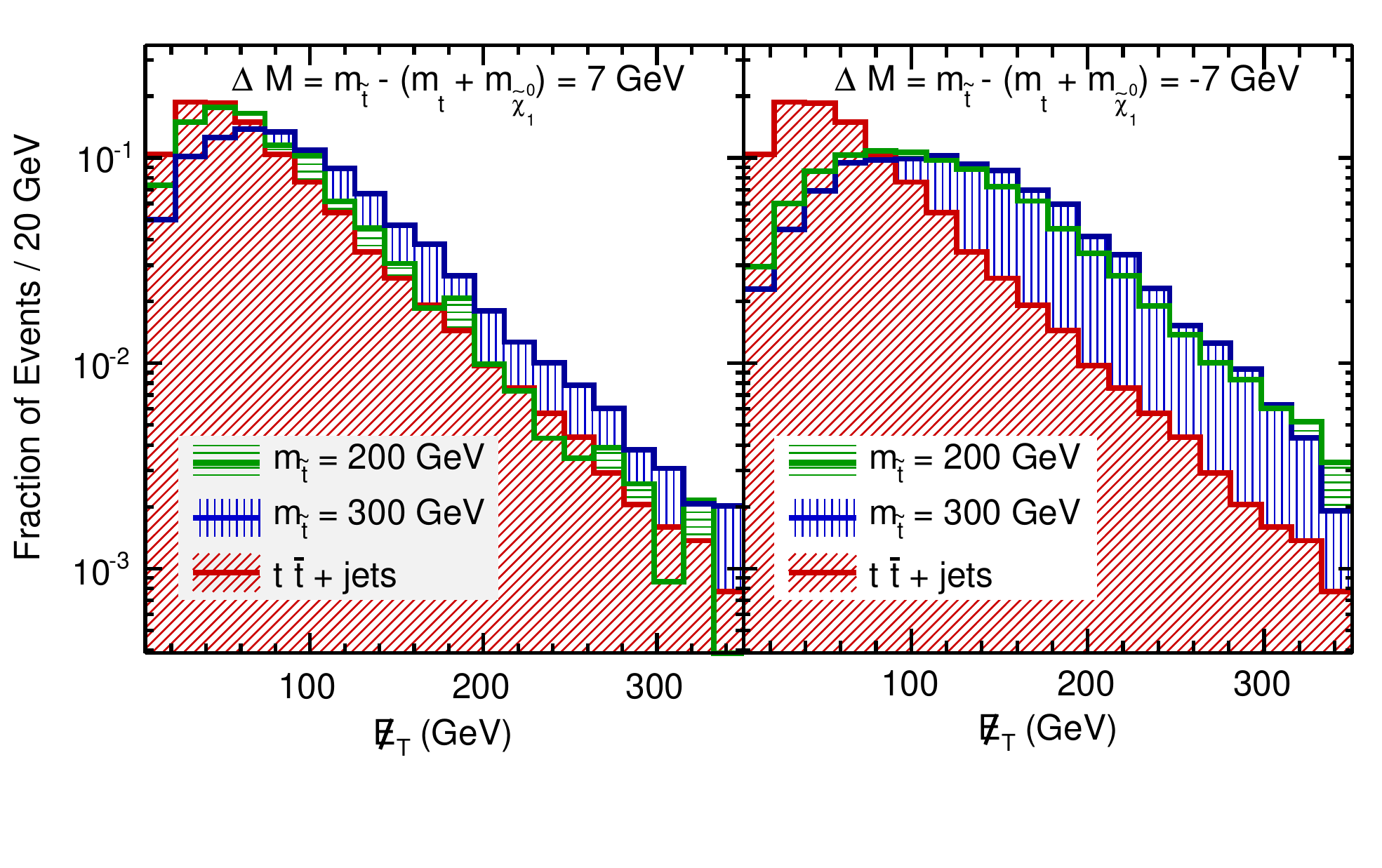}
\caption{Distributions of $\met$ normalized to unity for signal (green horizontally dashed histogram) and $\ttbar + $jets background (red diagonally dashed histogram) after VBF selections and lepton and $b$-jet requirements for the benchmark point with $m_{\st} = 400$ GeV, $m_{\neu{1}} = 220$ GeV.}
\label{MET_Norm1}
\end{figure}

Figure \ref{MET_Norm1} shows the distributions of $\met$ normalized to unity for signal (green horizontally dashed histogram) and $\ttbar + $jets background (red diagonally dashed histogram) after VBF selections and lepton and $b$-jet requirements for two benchmark points $\Delta M=7$ GeV and $\Delta M=-7$ GeV. From the figure, it is clear that the signal shows up as a broad enhancement in the tail of the $\met$ distribution.


  

It is clear from Figure 1 that there is significant benefit from prusuing a shape based analysis using the $\met$ distribution and the shape of the $\met$ distribution shows difference between $\Delta M=7$ and $\Delta M=-7$ GeV.  We indeed propose such a strategy and those results will be presented. 

We can calculate the significances $S/\sqrt{S+B}$, where $S$ and $B$ are the signal and background yields respectively, using a simple cut and count approach with the $\met$ preselections used in Tables I and II, keeping  $\Delta M = \pm7$ GeV, for various values of integrated luminosity at LHC14. We find that for $\Delta M=$7 GeV $m_{\st} \, \sim \, 390$ GeV ($320$ GeV) can be probed at 3$\sigma$ (5$\sigma$) level with 300 fb$^{-1}$ of integrated luminosity. The reach increases  to 560 GeV (470 GeV) at 3$\sigma$ (5$\sigma$) for 3000 fb$^{-1}$ of luminosity. For the three body case with $\Delta M = \, -7$ GeV, the reach for $\st$ is 320 (275) GeV at 3$\sigma$ (5$\sigma$) with 300 fb$^{-1}$ and  440 (380) GeV at 3$\sigma$ (5$\sigma$) with $3000$ fb$^{-1}$ integrated luminosity. 



\begin{figure}[!htp]
\centering
\includegraphics[width=3.0in]{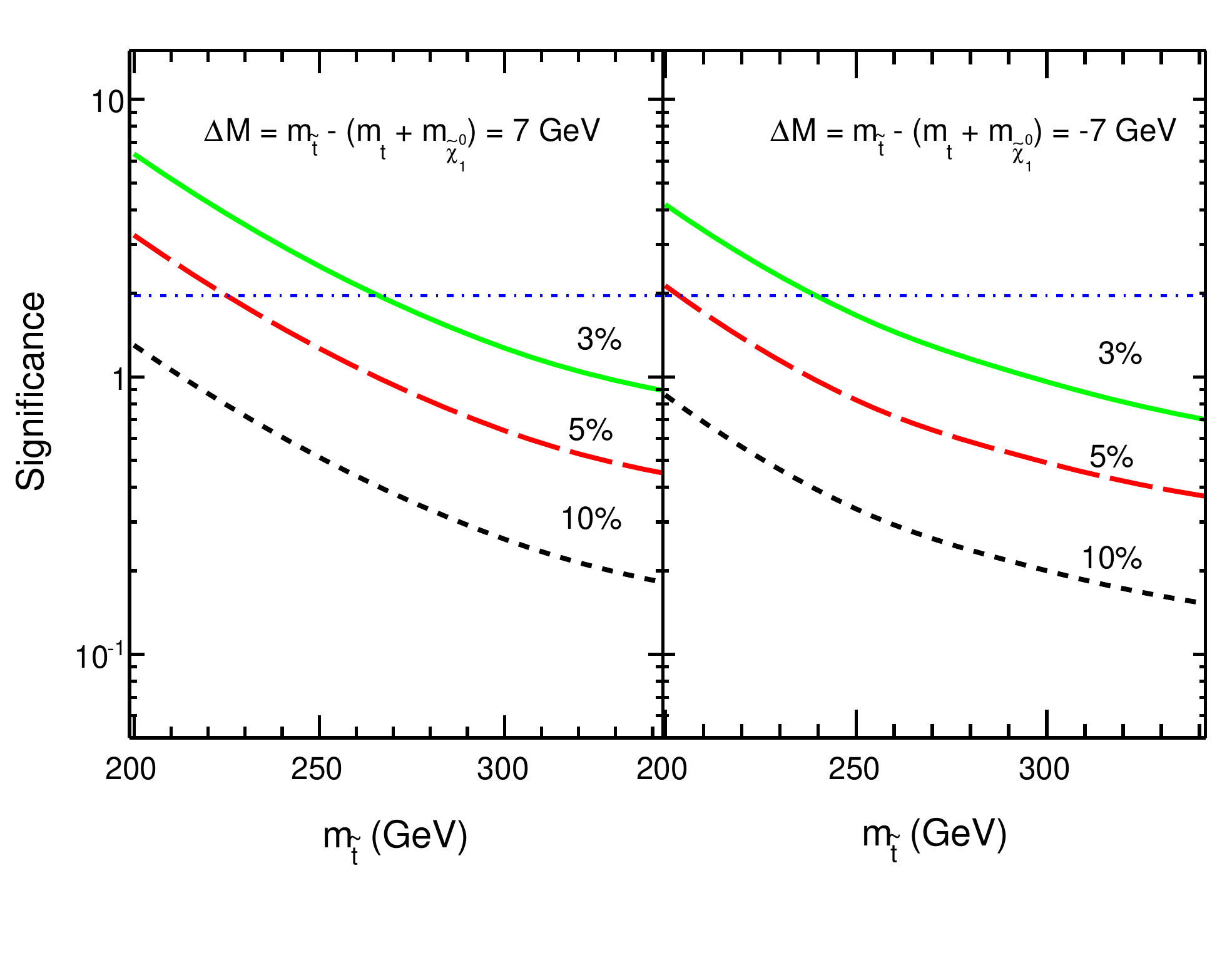}
\caption{Significance as a function of $m_{\st}$ for the $\Delta M=\pm7$ GeV for 3, 5 and 10\% systematics with integrated luminosities of $300$ fb$^{-1}$ at LHC14. The horizontal dotted line indicates 1.96 $\sigma$
or 95\% CL exclusion.}
\label{Stop_Three_Body_significance}
\end{figure}

{\it {\bf Systematics -}} The signal sensitivity considered thus far does not consider any source of systematic uncertainty.
In Fig.\ref{Stop_Three_Body_significance} we show signal significances under the consideration of of 3, 5 or 10\% for $\Delta M=\pm7$ GeV. The shape based analysis of the $\met$ distribution is performed using a 
binned likelihood following the test statistic based on the profile likelihood ratio.
The systematic uncertainties are incorporated via nuisance parameters following the frequentist approach. A local p-value is calculated as the probability under a background only hypothesis to obtain a value of the test statistic as large as that obtained with a signal plus background hypothesis. The significance $z$ is then determined as the value at which the integral of a Gaussian between $z$ and $\infty$ results in a value equal to the local p-value.
We find that the significance for $m_{\st}$ = 200 GeV and $\Delta M=-7$ GeV is expected to be 4(2)$\sigma $ for 300 fb$^{-1}$ luminosity with 3(5)\% systematic uncertainty, while the significance becomes 6(3)$\sigma$ for  $\Delta M = 7$ GeV. 

{\it {\bf Discussion -}} The main result of this paper is that the VBF topology can provide a feasible strategy to search for compressed top squarks. 
A major improvement over non-VBF searches in the compressed scenario is the efficacy of the $\met$ cut, due to the fact that top squarks are 
indirectly produced (e.g. by weak bosons, gluons, squarks, etc.) with a pair of high $E_{\rm T}$ tagging jets. We note that in the stealthy scenario, the $\neu{1}$ becomes massless, and the $\met$ cut loses its efficacy. We find that for $\Delta M=$7 GeV $m_{\st} \, \sim \, 390$ GeV ($320$ GeV) can be probed at 3$\sigma$ (5$\sigma$) level with 300 fb$^{-1}$ of integrated luminosity using  $S/\sqrt{S+B}$.  For the three body case  with $\Delta M = \, -7$ GeV, the reach for $\st$ is 320 (275) GeV at 3$\sigma$ (5$\sigma$) with 300 fb$^{-1}$. 
The significance gets degraded when systematic uncertainties are
taken into account. The significance  for $m_{\st}$ = 200 GeV and $\Delta M=7$ GeV is expected to be 6(3)$\sigma $ for 300 fb$^{-1}$ luminosity with 3(5)\% systematic uncertainty, while the significance becomes 4(2)$\sigma$ for the same stop mass with $\Delta M = -7$ GeV. There are no constraints for this parameter space point from the present data nor the ATLAS and CMS projections for the upcoming run. We also note that  the shape of the $\met$ distribution shows difference between $\Delta M=7$ and $\Delta M=-7$ GeV.
The determination of the systematic uncertainties due to the high pile-up conditions of the future is beyond the scope of this paper. It must be revisited with the expected performance of the upgraded ATLAS and CMS detectors.

{\it {\bf Acknowledgements -}} We would like to thank Howard Baer, Matt Buckley, Yu Gao, Tao Han, Kyoungchul Kong, Sanjay Padhi, and Lian-Tao Wang for helpful discussions. This work is supported in part by DOE Grant No.  DE-SC0010813, NSF Award PHY-1206044, and by the World Class University (WCU) project through the National Research Foundation (NRF) of Korea funded by the Ministry of Education, Science, and Technology (Grant No. R32-2008-000-20001-0). T.K. was also supported in part by Qatar National Research Fund under project NPRP 5 - 464 - 1 - 080. K.S. is supported by NASA Astrophysics Theory Grant NNH12ZDA001N.

\end{document}